\documentclass[pra,twocolumn,showpacs,superscriptaddress]{revtex4}

\usepackage{amsmath,amsfonts}
\usepackage[colorlinks]{hyperref}
\usepackage{graphicx}

\begin{document}

\title{Dynamics and stability of Bose-Einstein solitons in tilted optical
lattices}

\author{E.\ D\'{\i}az}
\affiliation{GISC, Departamento de F\'{\i}sica de Materiales, Universidad
Complutense, E-28040 Madrid, Spain}

\author{C.\ Gaul}
\affiliation{Physikalisches Institut, Universit\"{a}t Bayreuth,
D-95440 Bayreuth, Germany}

\author{R.\ P.\ A.\ Lima} 
\affiliation{GISC, Departamento de F\'{\i}sica de Materiales, Universidad
Complutense, E-28040 Madrid, Spain}
\affiliation{Instituto de F\'{\i}sica, Universidade Federal de Alagoas,
Macei\'{o} AL 57072-970, Brazil}

\author{F.\ Dom\'{\i}nguez-Adame}
\affiliation{GISC, Departamento de F\'{\i}sica de Materiales, Universidad
Complutense, E-28040 Madrid, Spain}

\author{C.\ A.\ M\"{u}ller}
\affiliation{Physikalisches Institut, Universit\"{a}t Bayreuth,
D-95440 Bayreuth, Germany}
\affiliation{Centre for Quantum Technolgies, National University of Singapore, Singapore 117653, Singapore}

\begin{abstract}
Bloch oscillations of Bose-Einstein condensates realize sensitive 
matter-wave interferometers. We investigate the dynamics and stability
of bright-soliton wave packets in one-dimensional tilted optical lattices with 
a modulated mean-field interaction $g(t)$.  
By means of a time-reversal argument, we prove the stability of
Bloch oscillations of breathing solitons that would be quasistatically
unstable. Floquet theory shows that these breathing solitons can be more
stable against certain experimental perturbations 
than rigid solitons or even noninteracting wave packets. 
\end{abstract}


\maketitle

Matter-wave interferometers are established as standard tools for
precision measurements of small forces. The tiny wavelength of ultracold atoms is an asset, and using Bose-Einstein condensates (BECs) can greatly enhance signal-to-noise ratios~\cite{Bouyer1997,Tuchman2009}. 
In this context, Bloch oscillations (BOs) of wave packets in tilted periodic potentials have been proven very useful~\cite{Clade2009a,Muller2009,Haller2009,Gustavsson2008a}. 
Much after their early prediction by Zener \cite{Bloch1929,Zener1934},
BOs were observed with electrons in semiconductor superlattices~\cite{Feldmann1992,Leo1992}, with cold atoms in optical
lattices~\cite{Ben1996,Anderson1998}, and with photons in waveguide
arrays \cite{Pertsch1999,Morandotti1999}.
BOs are very sensitive to dephasing
since they rely on coherent Bragg scattering of $k$-vectors
from one boundary of the Brillouin zone to the other.
The slightest lattice imperfection or interaction causes random scattering
of different $k$-components of a wave packet, thus broadening its momentum distribution and destroying coherent oscillations in real space. 

Long-living BOs of up to $10^4$ cycles with period $T_\text{B}$~\cite{Gustavsson2008} were achieved in a BEC experiment by tuning the scattering length between Cs atoms to zero with the help of a suitable Feshbach resonance~\cite{Donley2001,Koehler2006}. 
At finite interaction, it appears recommendable to use stable localized wave
packets, namely, soliton solutions to the nonlinear Schr\"{o}dinger equation, in
order to minimize the detrimental effects of interaction. Bright solitons arise
from a dispersion that counteracts the effect of the nonlinearity. As a rule, a
soliton can only be stable if the effective mass and interaction parameter have
opposite signs. In the free-space case of positive mass, bright solitons are
realized with atoms that attract each other~\cite{PerezGarcia1998,Khaykovich2002,Strecker2002}.
In a lattice, the mass becomes negative close to the edge of the Brillouin zone,
and there solitons can be prepared with repulsive
interaction~\cite{Eiermann2004}. During BO cycles,
the mass $m(t)$ changes its sign periodically. Thus, the interaction parameter $g(t)$ has to change accordingly
in order to respect the criterion of opposite signs; both spatial and temporal
nonlinearity management schemes to this effect have been proposed within the
framework of the Gross--Pitaevskii equation~\cite{Salerno2008,Bludov2009}.   

Independently, we have investigated the stability of BOs with an interaction
parameter $g(t)$ that is modulated harmonically in time, and we have found a whole family of
cases that yield stable BOs of breathing wave packets~\cite{Gaul2009}.
Conversely, many other instances of $g(t)$ result in rapid destruction of BOs.
Remarkably, the stability criterion developed in Ref.\ \cite{Gaul2009} is
obviously incompatible with the simple sign rule mentioned above. For instance, the Bloch-periodic modulations
$g(t) = \pm g_0 \cos(2\pi t/T_\text{B} )$ with opposite signs lead 
both to stable BOs, and this for the same mass $m(t)$. This prompts a question
that is as fundamentally interesting as it is important for practical
applications: Is solitonic stability  helpful to sustain BOs in general? 

In this work we show that stability of BOs is not conditioned on
wave-packet rigidity. 
We first introduce a bounded time in the
equation of motion that predicts perfectly stable breathing  
solutions, in full agreement with previous results \cite{Gaul2009}.
But even under these premises, rigid solitons could be expected to be more
robust against experimental imperfections. We study in detail the
relevant case of a BO decay caused by magnetic fields that oscillate off phase.
Contrary to expectations, we find that the breathing wave packet is more stable than the rigid soliton. 
For the experimentally realistic parameters chosen, the rigid soliton is very close to noninteracting. Thus, a harmonic
modulation of finite interaction turns out to effectively stabilize BOs.

In the mean-field regime, the BEC is described by the complex order parameter $\Psi(x,t)$ in a 1D optical lattice potential $V(x)=V_0 \cos^2(\pi x/b)$ with spacing~$b$. 
If the lattice is sufficiently deep, one may use a tight-binding
approximation, where the condensate is represented by a single complex number $\Psi_n(t)$ at each lattice site~\cite{Trombettoni2001}. Our starting point is thus the nonlinear equation of motion 
\begin{equation}\label{eqTightBinding}
i \hbar \dot \Psi_n = -J(\Psi_{n+1}+\Psi_{n-1}) + 
F b n  \Psi_{n} + g(t) |\Psi_{n}|^2 \Psi_{n}
\end{equation}
for the order parameter with normalization $\sum_i|\Psi_i|^2=1$.
Nearest-neighbor sites are coupled by the tunneling element  $J /E\approx 4
(V_0/E)^{3/4} \exp(-2\sqrt{V_0/E}) / \sqrt{\pi}$, where $E = \hbar^2
\pi^2/2mb^2$ is the lattice recoil energy \cite{Morsch2006}.
The interaction parameter $g = \sqrt{2{\omega_x\omega_y\omega_z \hbar m }/\pi} N a /J$ is derived from the scattering length $a$, the atom number $N$, and the local ground state determined by the lattice frequencies $\omega_i$ \cite{Trombettoni2001}.
The dispersion relation of this single-band model reads $\epsilon(p) = -2J
\cos(p b /\hbar)$. Its curvature or inverse mass $m^{-1} = 2 J b^2 \cos(p
b/\hbar)/\hbar^2$ determines the wave-packet dynamics under the
influence of a constant force $F$ that stems from a uniform acceleration of the BEC by, e.g.,
gravity.  
At fixed lattice geometry, the interaction parameter $g(t)$ can be controlled by external magnetic fields using suitable Feshbach resonances.

In the following, we use $J$ and $b$ as units of energy and length,
respectively, and set $\hbar=1$.   We tackle Eq.~\eqref{eqTightBinding} by
separating the rapidly varying Bloch phase $p(t)n$ from a  smooth envelope
$A(z,t)$ comoving with the center of mass $x(t)$:  $\Psi_n(t) = e^{i p(t) n}
A(n-x(t),t) e^{i \phi(t)} $. With $p(t) = - Ft$, $x(t) = x_0 + 2 \cos(Ft)/F$, and $\phi = \phi_0+ 2\sin(Ft)/F$, the envelope is found to obey the equation 
\begin{equation}
\label{eqAmplitudeEq}
 i \partial_t A = -\,\frac{1}{2m(t)} \, \partial_z^2 A + g(t) |A|^2 A \, ,
\end{equation}
neglecting higher spatial derivatives of $A$.  Note that choosing an
immobile wave packet with $p(0)=0$ as initial condition, fixes the
phase for the subsequent BOs.  The inverse mass $m(t)^{-1} = 2\cos(F
t)$ oscillates rapidly, but if the interaction is tuned such that
$m(t)g(t)<0$ varies slowly enough, Eq.~\eqref{eqAmplitudeEq} admits a
\emph{soliton} solution
\begin{equation}
\label{eqSoliton.mod}
A(z,t) = \frac{1}{\sqrt{2 \xi}}\,\frac{1}{\cosh\left(z/\xi\right)} \,
e^{-i \omega t} \, , 
\end{equation}
whose quasistatic width is $\xi = - {2}/[g(t)m(t)]>0$. To be stable, the soliton
must be able to follow this width adiabatically. 
Otherwise its breathing mode will be driven, and other excitations may be created.  Therefore, the least disruptive way
of accelerating a soliton of width $\xi_0$ is to impose a perfectly rigid
envelope by choosing $g_\text{r}(t) = -2 m(t)^{-1}/\xi_0 = g_\text{r} \cos(F t)$
with $g_\text{r} = -4/\xi_0<0$. More extensive studies based on this idea  have
been put forward in Refs.~\cite{Salerno2008,Bludov2009}.

\begin{figure}
\includegraphics[angle=-90,width=\linewidth]{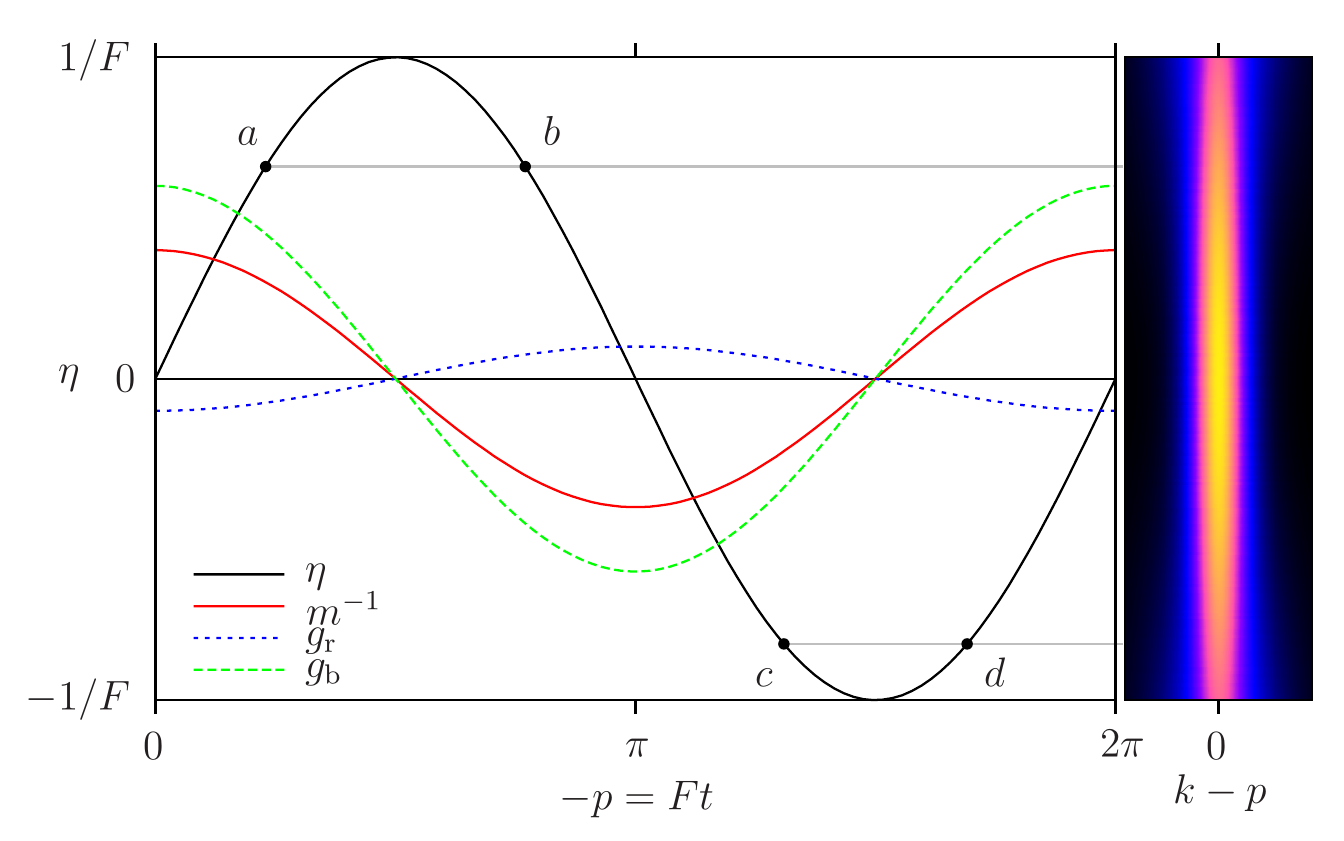}
\caption{Left panel: Time evolution scheme of stable BOs. The inverse mass
$m^{-1}$, the interaction parameter of the rigid soliton $g_\text{r}$ as well as
that of a breathing soliton $g_\text{b}$ are shown together with the bounded
time $\eta=\sin(Ft)/F$ as function of time or momentum $-p = F t$.  Right panel:
The $k$-space density [obtained by numerical integration of Eq.~\eqref{eqTightBinding}] 
of a breathing wave packet is a function of $\eta$ and
thus strictly periodic in $t$: the points in time $a$ and $b$ as well as
$c$ and $d$ show the same distribution, respectively.}
\label{figBreathingk}
\end{figure}

Let us contrast this reasoning with an analytical time-reversal argument that we
developed after studying the stability of BOs under harmonic modulations of
$g$~\cite{Gaul2009}. Quite generally, a rigid wave packet 
is by no means
necessary for persistent BOs. Already for the case $g(t) = 0$, breathing is the rule: 
In the first quarter of the Bloch cycle, the mass is positive and the wave packet spreads.
When the mass changes sign, the time evolution is reversed and the wave packet recovers its original shape at the edge of the Brillouin zone. 
Thus, the wave packet shows
periodic breathing on top of the BO, independently of its initial
shape. Also in the interacting case, one can find nontrivial functions
$g(t)$ in Eq.~\eqref{eqAmplitudeEq} compatible with this time reversal. 
Consider the class of 
periodic functions 
\begin{equation}
\label{eqStable2}
g(t) = \cos(F t)  P\big(\sin(F t) /F\big)\, ,
\end{equation}
where a factor $\cos(F t)$ can be separated from a polynomial
$P(\eta)$ in the bounded time variable 
\begin{equation}\label{eqBoundedTime}
\eta(t)=\frac{1}{2}\int_0^t m(s)^{-1}\mathrm{d}s = \frac{\sin(Ft)}{F} \, .
\end{equation} 
Because $\partial_t\eta(t)=[2m(t)]^{-1}$,
the explicit time dependence of the mass factorizes from all terms in
the equation of motion \eqref{eqAmplitudeEq} for $A(z,t)=\tilde A(z,\eta(t))$: 
\begin{align}
\label{eqAmplitudeEqBoundedTime}
 i \partial_\eta \tilde A(z,\eta) = & - \partial_z^2 \tilde A(z,\eta) + 
 P(\eta) |\tilde A|^2 \tilde A(z,\eta). 
\end{align}
The ensuing dynamics for $\tilde A(z,\eta)$ as function of $\eta$ may
be quite complicated. However, as $\eta(t)$ itself is a \emph{periodic
  function of time}, the solution $A(z,t)$ must also be periodic: Any
dynamics taking place in the first quarter of the Bloch period, while
$\eta$ runs from $0$ to $1/F$, is exactly reversed in the next
quarter, when $\eta$ runs back.  Figure~\ref{figBreathingk}
illustrates this argument by showing the time dependence of several
key quantities over one Bloch cycle, as well as a $k$-space density
plot with clearly visible breathing
dynamics~\footnote{Eq.~\eqref{eqStable2} covers all stable cases that
  are Bloch periodic, i.e.\ $\nu_2 = 1$ in Eq.\ (8) of
  \cite{Gaul2009}.  The cases $\nu_2 \neq 1$ are also covered by
  generalizing the bounded-time argument \cite{GaulPhD}.}.

We stress that Eq.~\eqref{eqStable2} includes both cases $g(t) = \pm
g_0 \cos(F t)$.  Although the $+{\cos}$ case does not fulfill the
soliton stability criterion $m(t) g(t) < 0$, the preceding time-reversal argument ensures that both cases lead to undamped Bloch
oscillations---at least within the approximations underlying the
equation of motion~\eqref{eqAmplitudeEq}.  As shown in the right-hand panel of 
Fig.~\ref{figBreathingk}, this prediction is confirmed by
numerical integration of the tight-binding model
\eqref{eqTightBinding} with a standard fourth-order Runge-Kutta
method.

Which of these stable solutions are the  most robust under variations of
experimental control parameters?  Indeed, even cold-atom experiments
suffer from slight imperfections, such as residual uncertainties in
the magnetic field controlling the interaction term $g(t)$.  For
instance, in the Innsbruck experiment \cite{Gustavsson2008} the
magnetic field is controlled up to $1\,$mG.  The slope of $61 a_0/{\rm
  G}$ at the zero of the Feshbach resonance turns this into an
uncertainty $\Delta a = 0.06 a_0$ in the scattering length, which is
converted to the uncertainty of the dimensionless tight-binding
interaction $\Delta g \approx 0.4$.
%
%
Note that this uncertainty $\Delta g$ is 
larger than the interaction 
$g_\text{r} = -4/\xi_0$ needed to create a rigid soliton
of only moderate width $\xi_0\gtrsim 10$. 
From this point of view, realizing a wide
rigid soliton is practically equivalent to switching the 
interaction off altogether.

\begin{figure}
\includegraphics[angle=-90,width=0.95\linewidth]{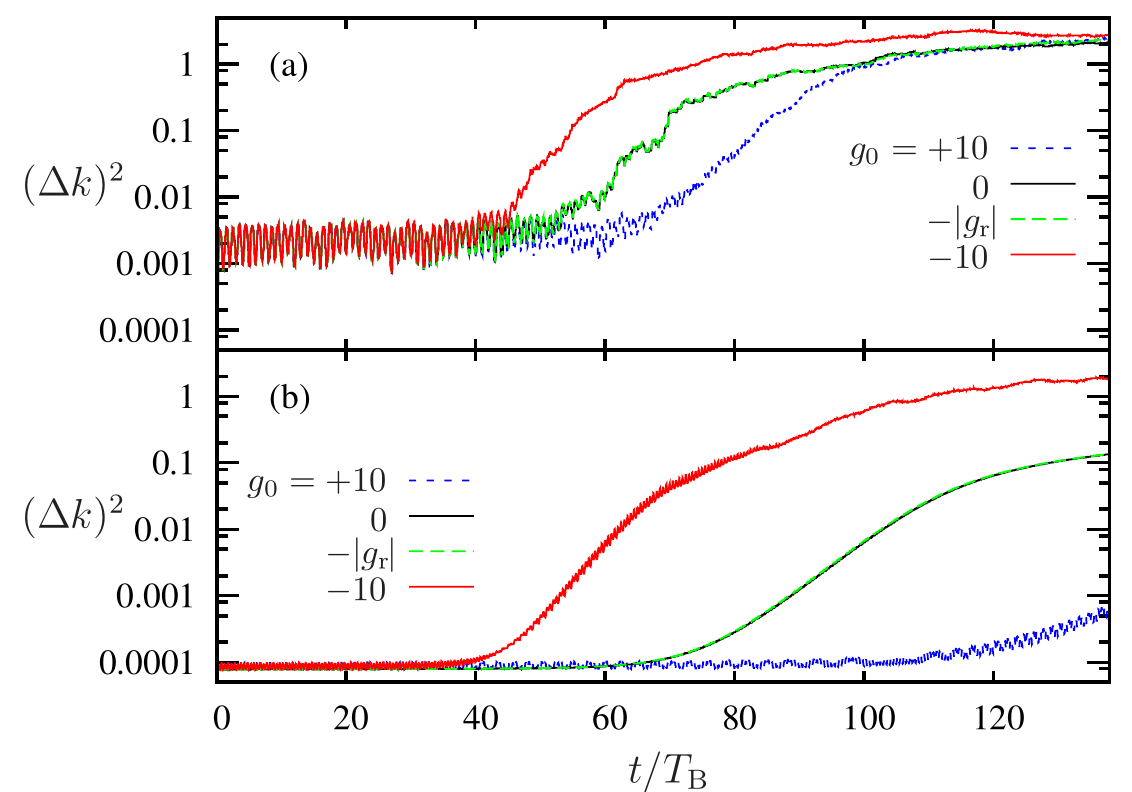}
\caption{Momentum width $(\Delta k)^2$ of a wave packet, Eq.~\eqref{eqSoliton.mod}, with initial
spatial width $\xi = 66.16$ 
(plus a small seed noise of $10^{-3}$ mimicking experimental inhomogeneities), 
Bloch-oscillating in a tilted
lattice with $F=2\pi/T_\text{B}=0.15$.
The interaction parameter is modulated as $g(t) = g_0 \cos(F t) + \nu(t)$.
(a) Random perturbation with frequencies $\Omega_n = n F /N$ below and above the Bloch frequency
$\nu(t) = \sum_{n=1}^{N^2} {\rm Re}[a_n \exp(i \Omega_n t) ]$, $N=5$.
The $a_n$ are complex random numbers with $\overline{a_n}=0$,
$\overline{|a_n|^2}=0.5$.
(b) Sine perturbation $\nu(t) = 0.5 \sin(F t)$. 
In all cases the lifetime increases with the perturbation amplitude $g_0$, from the antibreathing wave packet $g_0 = -10$, 
over the rigid soliton $g_0 = g_\text{r} = -0.06$ 
and the linear wave packet $g_0=0$ 
to the breathing soliton $g_0 = +10$. 
}
\label{figPhase}
\end{figure}

We study numerically the effect of perturbations of $g(t)$ by numerical integration of Eq.~\eqref{eqTightBinding}. 
Rather than the strong force $F \approx 34$ in a vertical lattice \cite{Gustavsson2008}, 
we choose a smaller force $F=0.15$, corresponding to a slighter tilt. 
This results in a longer Bloch period and a higher sensitivity to dephasing.
The broadening of the momentum distribution is directly measurable
from the experimental time-of-flight images and signals the decay of
the wave packet and destruction of BOs in real space.  In
Fig.~\ref{figPhase} the $k$-space variance $(\Delta k)^2$ is shown for
$g(t) = g_0 \cos(F t) + \nu(t)$ with two different types of
perturbations $\nu(t)$ and different modulation strengths $g_0$.  In
all cases, the momentum distribution starts to broaden at a certain
time.  At a given perturbation, the rigid soliton and the linear wave
packet, show greater resilience than the strongly antibreathing wave
packet, but surprisingly the breathing wave packet survives even
longer. Obviously, the +cos modulation stabilizes the Bloch
oscillations against uncontrolled variations of the 
interaction parameter $g$.

\begin{figure}
\includegraphics[angle=-90,width=\linewidth]{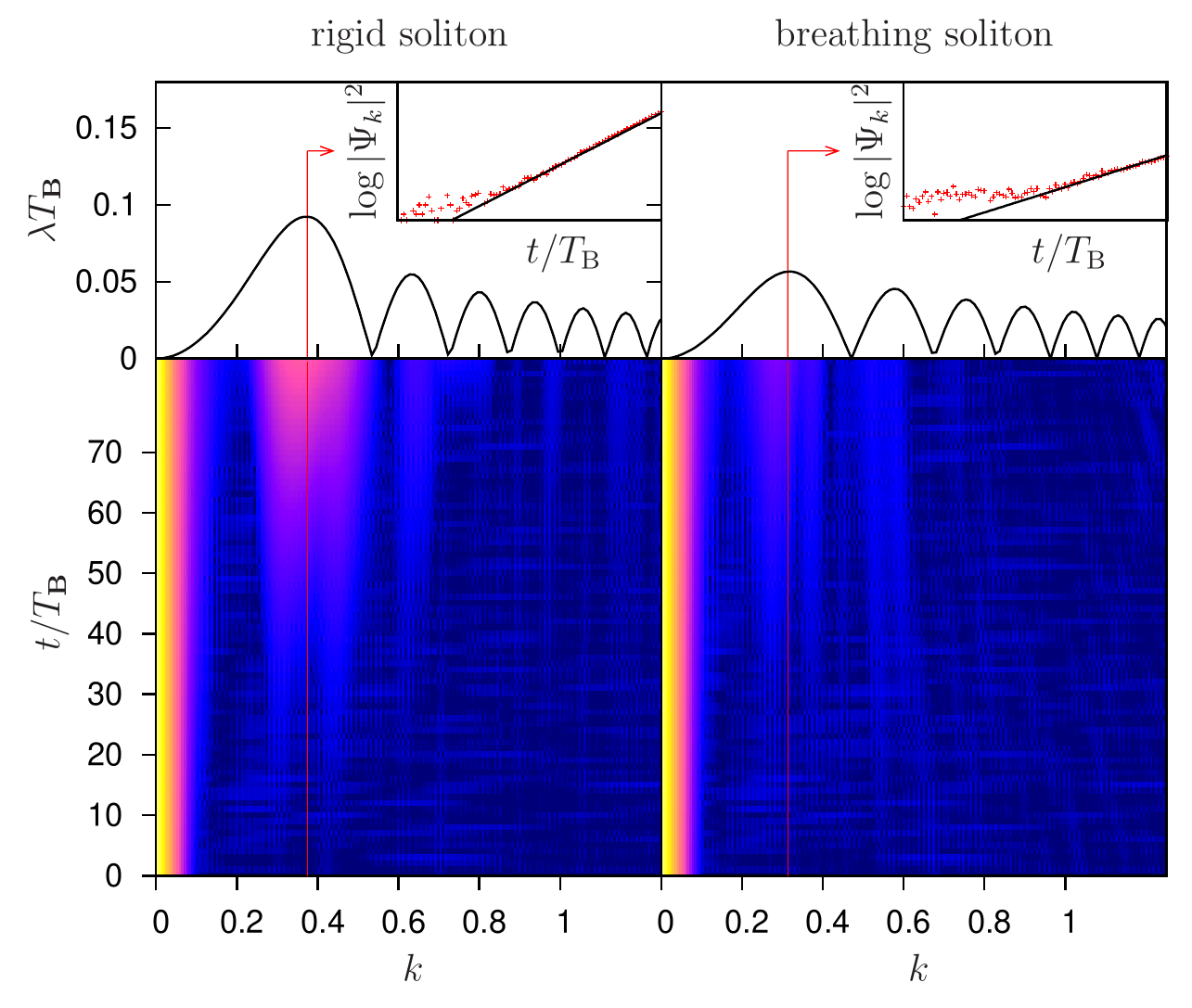}
\caption{Decay of a rigid (left) and a breathing (right) soliton under the harmonic perturbation of Fig.~\ref{figPhase}~(b).
Upper panel: linear stability growth rate, Eq.\ \eqref{eqLyapunov}. 
Lower panel: Stroboscopic plot of the $k$-space density on a logarithmic color scale. 
Inset: Location and growth of the most unstable
mode agree with the analytical prediction.
}\label{figKGrowth}
\end{figure}

Comparing a random superposition of different frequencies [Fig.\ \ref{figPhase}(a)] to a  
Bloch-periodic off-phase perturbation 
proportional to $\sin(F t)$ [Fig.\ \ref{figPhase}(b)], we trace back 
the differences in lifetime to different sensitivities to the sine perturbation.
In the following, we thus consider $g(t) = g_0\cos(Ft)+g_1\sin(Ft)$ with an
off-phase perturbation with amplitude $g_1$ of order $\Delta g$. 
In order to understand quantitatively why the $+{\cos}$-modulated wave
packet can be more robust than the rigid soliton, we perform a
homogeneous stability analysis. The sudden growth of the momentum
variance suggests an instability due to the creation of small
fluctuations. If these perturbations occur on a length scale much
shorter than the size of the wave packet, the wave packet can be taken to be
locally homogeneous, $|A(z,t)|^2 \approx n_0 = 1/2\xi$.  The real and
imaginary parts of the small fluctuations to linear order, $s$ and $d$,
respectively, obey equations of motion that decouple in Fourier
modes~\cite{Gaul2009}:
\begin{equation}\label{eqLinearStability} 
\begin{aligned}
\dot d_k &= -[k^2\cos(Ft) +2 n_0 g(t)] s_k, \\ 
\dot s_k &=  k^2\cos(Ft) d_k. 
\end{aligned}
\end{equation}
Thanks to their linearity and time periodicity, these equations allow
us to apply Floquet theory \cite{Teschl2008}.  Integrating
Eq.~\eqref{eqLinearStability} over a single period determines 
the Lyapunov exponent $\lambda_k$ that characterizes the exponential
growth of mode $k$.  For the present case $g(t) = g_0 \cos(F t) + g_1
\sin(F t)$, Eq.~\eqref{eqLinearStability} can be solved analytically
for $g_1=0$, using a Bogoliubov transformation $\gamma_k =
\sqrt{\omega_k/k^2} s_k + i \sqrt{k^2/\omega_k} d_k$, with $\omega_k =
\sqrt{k^2 (k^2 + 2 n_0 g_0)}$.  Perturbation theory to first order in
$g_1$ then predicts the growth of mode $\gamma_k$ with
\begin{align}\label{eqLyapunov}
 \lambda_k =  k^2\left| \frac{g_1 n_0}{\omega_k} J_1(2\omega_k/F) \right| ,
\end{align}
where $J_1$ is the Bessel function of the first kind.
For the two cases of primary interest---the rigid and the breathing soliton---the 
upper panel of Fig.~\ref{figKGrowth} shows this Lyapunov exponent,
indistinguishable from the value obtained by numerical solution of
Eq.~\eqref{eqLinearStability}, as function of $k$. 
Mainly the prefactor $k^2/|\omega_k|=|1+2g_0n_0/k^2|^{-1}$ makes  
the Lyapunov exponents of the breathing wave packet ($g_0=10$) smaller
than those of the rigid soliton ($g_0=g_\text{r}=-0.06$).  
\footnote{For $g_0<0$, very small $k$-values such that $k^2<2n_0|g_0|$
have imaginary Bogoliubov frequencies $\omega_k$, but
Eq.~\eqref{eqLyapunov} allows for a smooth analytic continuation,
cf.~Fig.~\ref{figKGrowth}.}

The Lyapunov exponent \eqref{eqLyapunov} provides a rather faithful
portrait of the $k$-space evolution as obtained by the numerics,
plotted in the lower panels of Fig.~\ref{figKGrowth}. Notably,
excitations grow exclusively in the intervals with the largest
Lyapunov exponents, found at $k$-values such that $k^2\gg n_0g_0$.
Thus, the most unstable mode (indicated by the vertical line) is found
close to $k_*$ defined by $J_1'(2k_*^2/F)=0$, which gives $k_*\approx
0.96\sqrt{F}\approx 0.37$.  The predicted growth rate agrees very well
with the numerical data (inset in the upper panel).  These predictions
remain valid also for smaller $g_0$. In the limit $g_0\to0$, the
differences between different wave packets disappear, and we recover
the case of a pure $\sin(Ft)$-modulation that was analyzed in
\cite{Gaul2009}, and for which Eq.\ \eqref{eqLyapunov} now provides an
analytical expression.

In essence, this linear-stability analysis applies whenever the
excitations are well decoupled in $k$-space from the original wave
packet (the central peak around $k=0$ visible in
Fig.~\ref{figKGrowth}).  Note that other types of perturbations, e.g.,
a constant offset $g_2$ as in $g(t) =
g_0\cos(Ft)+g_2$~\cite{Gong2009}, can cause a more homogeneous
broadening of the $k$-space distribution.  Its effect is better
captured by an ansatz in terms of collective variables (cf.\
\cite{Gaul2009}).  A detailed analysis is beyond the scope of the
present work and will be addressed in a forthcoming publication.


In conclusion, we have connected the physics of solitons in lattices
with a stability analysis of BOs under harmonic variations of the
interaction.  Stability of BOs does not rely on soliton
stability.  Instead, the wave packet may start to fall apart, but
comes back by virtue of periodic time reversal.  In the presence of
instability-inducing perturbations, a modulation of the interaction
can make the wave packet more robust.  We explain this behavior
quantitatively via linear stability analysis within Floquet theory.
Finally, let us stress that these results play an important role in
the design of accurate and reliable matter-wave interferometers based
on BOs.


Travel between Bayreuth and Madrid was supported by
the joint program Acciones Integradas of 
DAAD  and 
MEC. C.G.\ and C.A.M.\  acknowledge financial
support from DFG, and
thank H.-C. N\"agerl and his group for hospitality and helpful
discussions. Work at Madrid was supported by MEC (Project
MOSAICO). Work at Macei\'o was partially supported by 
CNPq 
and 
CAPES.


\begin{thebibliography}{10}

\bibitem{Bouyer1997}
P.~Bouyer and M.~A.~Kasevich, {Phys. Rev. A} \textbf{56}, R1083
(1997).

\bibitem{Tuchman2009}
A.~K.~Tuchman and M.~A.~Kasevich, {Phys. Rev. Lett.} \textbf{103}, 130403
(2009).

\bibitem{Clade2009a}
P.~Clad\'{e}, S.~Guellati-Kh\'{e}lifa, F.~Nez, and F.~Biraben, {Phys. Rev.
  Lett.} \textbf{102}, 240402 (2009).

\bibitem{Muller2009}
H.~M\"{u}ller, S.~Chiow, S.~Herrmann, and S.~Chu, {Phys. Rev. Lett.}
  \textbf{102}, 240403 (2009).

\bibitem{Haller2009}
E.~Haller, \emph{et~al.}, {Science} \textbf{325}, 1224 (2009).

\bibitem{Gustavsson2008a}
M.~Gustavsson, \emph{et~al.}, arXiv:0812.4836 (2008).

\bibitem{Bloch1929}
F.~Bloch, {Zeitschr. Phys.} \textbf{52}, 555 (1929).

\bibitem{Zener1934}
C.~Zener, {Proc. R. Soc. Lond. A} \textbf{145}, 523 (1934).

\bibitem{Feldmann1992}
J.~Feldmann, \emph{et~al.}, {Phys. Rev. B} \textbf{46}, 7252 (1992).

\bibitem{Leo1992}
K.~Leo, P.~H. Bolivar, F.~Br{\"u}ggemann, R.~Schwedler, and K.~K{\"o}hler,
  {Sol. State Comm.} \textbf{84}, 943  (1992).

\bibitem{Ben1996}
M.~Ben~Dahan, E.~Peik, J.~Reichel, Y.~Castin, and C.~Salomon, {Phys. Rev.
  Lett.} \textbf{76}, 4508 (1996).

\bibitem{Anderson1998}
B.~P. Anderson and M.~A. Kasevich, {Science} \textbf{282}, 1686 (1998).

\bibitem{Pertsch1999}
T. Pertsch, \emph{et~al.}, {Phys. Rev. Lett.} \textbf{83}, 4752 (1999).

\bibitem{Morandotti1999}
R. Morandotti, \emph{et~al.}, {Phys. Rev. Lett.} \textbf{83}, 4756 (1999).

\bibitem{Gustavsson2008}
M.~Gustavsson, \emph{et~al.}, {Phys. Rev. Lett.} \textbf{100}, 080404
  (2008).

\bibitem{Donley2001}
 E.~A. Donley, \emph{et~al.}, {Nature} \textbf{412}, 295 (2001).

\bibitem{Koehler2006}
T.~K{\"o}hler, K.~G{\'o}ral, and P.~S. Julienne, {Rev. Mod. Phys.}
  \textbf{78}, 1311 (2006).

\bibitem{PerezGarcia1998}
V.~M.~P\'erez-Garc\'ia, H.~Michinel, and H.~Herrero, {Phys. Rev. A} \textbf{57}, 3837 (1998).

\bibitem{Khaykovich2002}
L.~Khaykovich, \emph{et~al.}, {Science} \textbf{296}, 1290 (2002).

\bibitem{Strecker2002}
K.~E. Strecker, G.~B. Partridge, A.~G. Truscott, and R.~G. Hulet, {Nature}
  \textbf{417}, 150 (2002).

\bibitem{Eiermann2004}
B.~Eiermann, \emph{et~al.}, {Phys. Rev. Lett.} \textbf{92}, 230401 (2004).

\bibitem{Salerno2008}
M.~Salerno, V.~V. Konotop, and Y.~V. Bludov, {Phys. Rev. Lett.}
  \textbf{101}, 030405 (2008).

\bibitem{Bludov2009}
Y.~V. Bludov, V.~V. Konotop, and M.~Salerno, {J. Phys. B: At. Mol. Opt.
  Phys.} \textbf{42}, 105302 (2009).

\bibitem{Gaul2009}
C.~Gaul, R.~P.~A. Lima, E.~D{\'i}az, C.~A. M{\"u}ller, and
  F.~Dom{\'i}nguez-Adame, {Phys. Rev. Lett.} \textbf{102}, 255303 (2009).

\bibitem{Trombettoni2001}
A.~Trombettoni and  A.~Smerzi, {Phys. Rev. Lett.} \textbf{86}, 2353  (2001).

\bibitem{Morsch2006}
O.~Morsch and M.~Oberthaler, {Rev. Mod. Phys.} \textbf{78}, 179 (2006).

\bibitem{GaulPhD}
C. Gaul, \href{http://opus.ub.uni-bayreuth.de/volltexte/2010/678}{Ph.D.\ thesis, Universit\"at Bayreuth, 2010}.

\newcommand{\enquote}[1]{``#1''}
 \bibitem{Teschl2008} 
N.G~ Markley, Principles of Differential Equations (Wiley,
Hoboken, NJ, 2004), Ch. 5.4.


\bibitem{Gong2009} 
J.~Gong, L.~Morales-Molina, and P.~H\"anggi, 
{Phys. Rev. Lett.} \textbf{103}, 133002 (2009).



\end{thebibliography}
\end{document}